\documentclass[twocolumn]{article}
\usepackage{graphicx} 
\usepackage{subcaption}
\usepackage{makecell}
\usepackage{caption}
\usepackage{amsmath}
\usepackage{xcolor}
\usepackage{authblk}

\begin{document}

\title{High Information Density and Low Coverage Data Storage in DNA with Efficient Channel Coding Schemes}

\author{
        Yi Ding\textsuperscript{1,2}
        , Xuan He\textsuperscript{1}*
        , Tuan Thanh Nguyen\textsuperscript{2}
        , Wentu Song\textsuperscript{2}
        , Zohar Yakhini\textsuperscript{3,4}
        , Eitan Yaakobi\textsuperscript{3}
        , Linqiang Pan\textsuperscript{5}
        , Xiaohu Tang\textsuperscript{1}
        , and Kui Cai\textsuperscript{2}*
        }
\maketitle

\footnotetext[1]{
Information Coding and Transmission Key Lab of Sichuan Province, Southwest Jiaotong University, Chengdu 611756, Sichuan, China.}

\footnotetext[2]{Science, Mathematics and Technology (SMT) Cluster, Singapore University of Technology and Design, Singapore 487372.}

\footnotetext[3]{Faculty of Computer Science, Technion - Israel Institute of Technology, Haifa 3200003, Israel.}

\footnotetext[4]{School of Computer Science, RUNI, Herzliya	4615200, Israel.}

\footnotetext[5]{Key Laboratory of Image Information Processing and Intelligent Control of Education Ministry of China, School of Artificial Intelligence and Automation, Huazhong University of Science and Technology, Wuhan 430074, Hubei, China.}
\renewcommand\thefootnote{}
\footnotetext{*Corresponding authors: Kui Cai (e-mail: cai\_kui@sutd.edu.sg) and Xuan He (e-mail: xhe@swjtu.edu.cn).}

\begin{abstract}
DNA-based data storage has been attracting significant attention due to its extremely high data storage density, low power consumption, and long duration compared to conventional data storage media. Despite the recent advancements in DNA data storage technology, significant challenges remain. In particular, various types of errors can occur during the processes of DNA synthesis, storage, and sequencing, including substitution errors, insertion errors, and deletion errors. Furthermore, the entire oligo may be lost. In this work, we report a DNA-based data storage architecture that incorporates efficient channel coding schemes, including different types of error-correcting codes (ECCs) and constrained codes, for both the inner coding and outer coding for the DNA data storage channel. We also carried out large scale experiments to validate our proposed DNA-based data storage architecture. Specifically, 1.61 and 1.69 MB data were encoded into 30,000 oligos each, with information densities of 1.731 and 1.815, respectively. It has been found that the stored information can be fully recovered without any error at average coverages of 4.5 and 6.0, respectively. This experiment achieved the highest net information density and lowest coverage among existing DNA-based data storage experiments (with standard DNA), with data recovery rates and coverage approaching theoretical optima.

\end{abstract}

\section{Introduction}





\begin{figure*}[!t]
    \centering
    \includegraphics[width=1\linewidth]{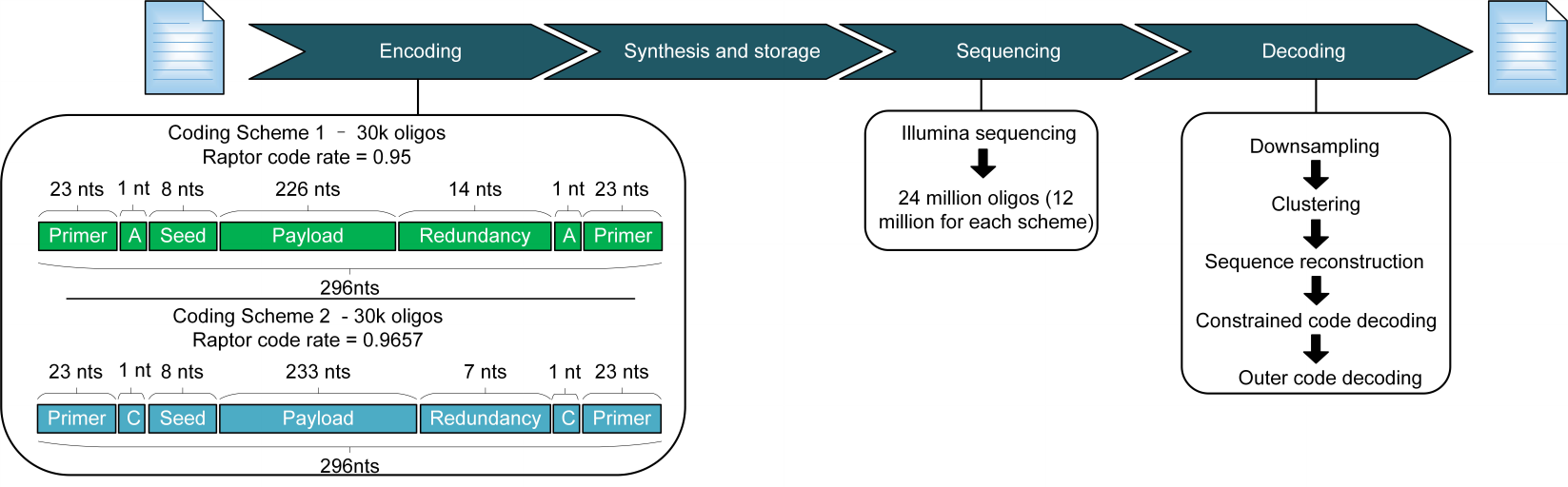}
    \caption{The workflow of the designed DNA-based data storage system with the structure of an oligo. In the experiment, 30,000 oligos of length 296 nucleotides were synthesized for two different coding schemes. In Coding Scheme 1, each oligo has 226 nucleotides as data payload, 8 nucleotides as seed, 14 nucleotides as redundancy to remove homopolymers runs of more than 4, and to detect or correct errors at DNA base level, 2 nucleotides as the indicator of different coding schemes, and 46 nucleotides as primers. In Coding Scheme 2, each oligo has 233 nucleotides as data payload, 8 nucleotides as seed, 7 nucleotides as redundancy to remove homopolymers runs of more than 4, and to detect or correct errors at DNA base level, 2 nucleotides as the indicator of different coding schemes, and 46 nucleotides as primers. After the processes of DNA synthesis and storage, Illumina sequencing was applied to the DNA pool to obtain 24 million sequenced oligos. The source data was recovered after the decoding processes.}
    \label{fig:workflow}
\end{figure*}

\begin{table*}[!h]
    \centering
    \caption{The comparison between the designed information storage architecture and previous works.}
    \begin{tabular}{|c|c|c|c|c|}
        \hline
         & \makecell[c]{Data size \\ (MB)} & \makecell[c]{Number \\ of \\ oligos} & \makecell[c]{Net \\ information \\ density \\ (bits/nt)} & Coverage(x) \\
         \hline
         Church et al. \cite{church2012next} & 0.65 & 54,898 & 0.83 & 3000 \\
         \hline
         Goldman et al. \cite{goldman2013towards} & 0.63 & 153,335 & 0.33 & 51 \\
         \hline
         Grass et al. \cite{grass2015robust} & 0.08 & 4,991 & 1.14 & 372 \\
         \hline
         Bornholt et al. \cite{bornholt2016dna} & 0.15 & 151,000 & 0.88 & 40 \\
         \hline
         Blawat et al. \cite{blawat2016forward} & 22 & 1,000,000 & 0.92 & 160 \\
         \hline
         Erlich et al. \cite{erlich2017dna} & 2.11 & 72,000 & 1.57 & 10.5 \\
         \hline
         Organick et al. \cite{organick2018random} & 200.2 & 13,400,000 & 1.1 & 5 \\
         \hline
         Wang et al. \cite{wang2019high} & 0.38 & 11,400 & 1.67 & 10 \\
         \hline
         This work (Coding Scheme 1) & 1.61 & 30,000 & 1.731 & 4.5 \\
         \hline
         This work (Coding Scheme 2) & 1.69 & 30,000 & 1.815 & 6.0 \\
         \hline
    \end{tabular}
    
    \label{tab:comparison}
\end{table*}

Digital data is currently generated by humanity at an exponentially increasing rate. It is predicted that the demand for data storage globally will surge to $1.75 \times 10 ^ {14}$ GB in 2025 \cite{reinsel2017data}. Conventional storage technologies can hardly meet future needs due to their inherent limitations, such as restricted information density and longevity. For example, the storage density of the most advanced enterprise solid state drive (SSD) is less than 1 GB/mm$^{3}$. Therefore, it requires a volume of $1.75 \times 10^5$ m$^3$ to store all the data that will be produced in 2025. The DNA-based data storage technology has emerged as a promising solution to the current data storage crisis, due to its extremely high data storage density, long lasting stability of hundreds to a thousand years, and ultra-low power consumption for operation and maintenance \cite{heckel2019acharacterization}. With a storage density of around $1.7 \times 10 ^ {10} $ GB/g \cite{organick2020probing}, only $0.01$ m$^3$ DNA is required to store the same amount of data predicted for 2025, thus demonstrating the significant potential of DNA-based data storage technology. 

The basic unit of DNA data storage is a DNA strand (or oligonucleotides, or oligo in short), which is a sequence of four types of nucleotides: adenine (A), cytosine (C), guanine (G), and thymine (T). Each nucleotide (also known as the DNA base) can represent two bits of information. \textbf{DNA data storage mainly involves data encoding, DNA synthesis (writing), storage, sequencing (reading), and data decoding processes.} As shown by Figure \ref{fig:workflow}, the input binary user data is first encoded into quaternary base sequences by the channel encoder. Modern array-based synthesizer synthesizes up to $10^6$ copies of each DNA strand of about 200 nucleotides in length, which is essentially the write process. All these oligos are then mixed and stored in a DNA oligo pool. During the reading process, a so-called polymerase chain reaction (PCR) is typically performed to amplify the oligos in the pool \cite{heckel2019acharacterization}. Reading information in DNA data storage is realized by randomly and independently sequencing the oligos in the pool, with each sequenced oligo as one read. The total number of reads is called sequencing depth or coverage. After channel decoding, the original information stored in the DNA strands will be recovered. 
A DNA strand/oligo can become lost or corrupted by insertion, deletion, or substitution errors during the synthesis, storage, and sequencing processes. Therefore, error-correcting codes (ECCs) are applied to detect or correct these errors by introducing certain amount of redundancy. In a typical DNA-based data storage system, an inner code is used to detect/correct the errors within a DNA strand, whereas an outer code is applied to recover the missing DNA strands and correct the erroneous strands from inner decoding. Furthermore, two biological DNA base patterns that significantly increase the occurrence of errors for most synthesis and sequencing techniques are the long homopolymer run (a segment of DNA strand with identical nucleotides), and unbalanced GC-content (deviation of the proportion of G and C from 50\%) within a DNA strand. This motivated the application of the constrained code to convert binary data into DNA strands which do not have long runs of homopolymer and whose GC-content is around 50\% \cite{immink2017design, nguyen2021capacity}. 

Several experiments with different DNA data storage architectures have been carried out to demonstrate the efficiency of DNA-based data storage systems. Except for \cite{church2012next}, most of these works have employed various channel coding techniques \cite{church2012next, goldman2013towards, grass2015robust, bornholt2016dna, blawat2016forward, erlich2017dna, organick2018random, wang2019high, welzel2023dna}. In particular, Reed-Solomon (RS) codes were adopted as the outer code in \cite{grass2015robust, blawat2016forward, organick2018random, wang2019high}, while RS codes and Cyclic Redundancy Check (CRC) codes were utilized as the inner code in \cite{grass2015robust, erlich2017dna, organick2018random} and \cite{blawat2016forward, wang2019high, welzel2023dna}, respectively. The repetition methods have also been employed in \cite{goldman2013towards, bornholt2016dna} to tackle the substitution and erasure errors.

In \cite{erlich2017dna},  Erlich et al. adopted the Luby transform (LT) codes, the first practical class of fountain codes \cite{mackay2005fountain}, as the outer code, highlighting that the rateless nature of fountain codes enables its efficient combination with the constrained code for DNA data storage. In particular, the LT encoder first generate a large number of codewords, and it then rejects those that do not satisfy the homopolymer runlength and the GC-content constraints. Later, another class of fountain codes, Raptor codes \cite{shokrollahi2006raptor} were employed as the outer code \cite{welzel2023dna}. That is, Welzel et al. \cite{welzel2023dna} proposed a concatenated coding scheme capable of generating oligos that adhere to user-defined constraints and correcting edit and erasure errors. 

In this work, we propose a DNA data storage architecture based on efficient channel coding techniques. The workflow of our proposed architecture is depicted in Figure \ref{fig:workflow}. Its performance was evaluated through experiments, and the corresponding main results and their comparison with previous experimental works are shown in Table \ref{tab:comparison}. To the best of our knowledge, this study achieves the highest information density and the lowest coverage compared to all previous experiments (with standard DNA).

\begin{figure*}[!t]
    \centering
    \includegraphics[width=1\linewidth]{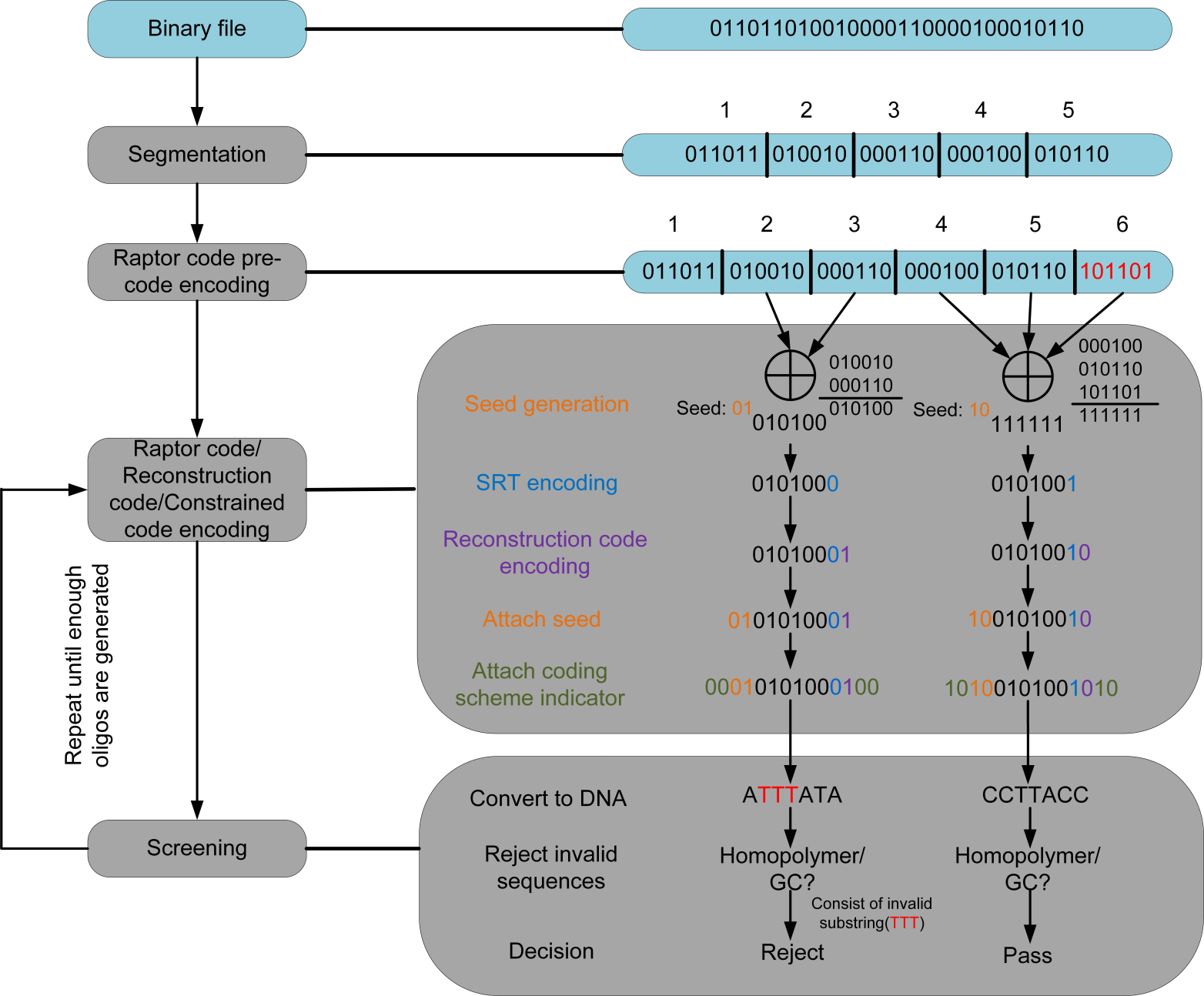}
    \caption{Encoding process of the designed DNA-based data storage architecture. For illustration purposes, we consider an input binary file of 30 bits, partitioned into 5 segments of 6 bits each. The seed is assumed to be 2 bits long. See Sections 4.1 to 4.3 for full details of each encoding step.}
    \label{fig:encoding}
\end{figure*}

\section{Results}

\subsection{Encoding Schemes}
\label{Sec:Encoding Schemes}
In our proposed DNA data storage architecture, the source data is first encoded by a combined encoder of the outer ECC code, inner ECC code, and the constrained code. The corresponding encoding process is illustrated by Figure \ref{fig:encoding}. We applied two coding schemes, indicated by two nucleotide variations in the oligo (see Figure \ref{fig:workflow}, left part): one positioned before the seed and the other after the redundancy. Both coding schemes follow the same encoding workflow as depicted in Figure \ref{fig:encoding}. Compared with Coding Scheme1, Coding Scheme 2 has a higher code rate and lower logical redundancy, thereby achieving a higher information density. However, due to its weaker error correction capability, the corresponding coverage for error free recovery of the input data is higher than that of Coding Scheme 1. In this work, we mapped nucleotides `A, T, C, G' to `00, 01, 10, 11' or `0, 1, 2, 3', respectively. We alternate between these notations when convenient. 
In the experiment, a modified Raptor 10 code (modified-R10, see Section \ref{Sec:R10}) \cite{luby2007raptor} and a single edit reconstruction code (see Section \ref{Sec:Method_Reconstruction}) were adopted as the outer ECC and the inner ECC, respectively. For the constrained code, we employed a modified sequence replacement technique (modified-SRT, see Section \ref{Sec:Method_SRT}) to remove long homopolymers runs, and leveraged the rateless nature of fountain codes to ensure that only the oligos with required homopolymers runs and GC-content were selected as the output of the encoder. 
As illustrated in Figure \ref{fig:encoding}, the encoding process proceeds as follows. First, the binary input data file was divided into non-overlapping segments of fixed length, with each segment representing a source symbol. Through a pre-encoding process using Raptor code (see Section \ref{Sec:R10}), this sequence of source symbols was encoded into a sequence of intermediate symbols, which were then further encoded by the LT encoder.

The resulting codeword consists of two parts: a unique seed and the data payload. In this work, the seed started at zero and increased by one with each encoding round until the encoder output reaches 30,000 oligos for both coding schemes. The seed initialized a pseudorandom number generator (PRNG), which was used to randomly select a subset of intermediate symbols. These selected symbols were combined bitwise under a binary field to form the data payload.

Next, the data payload was encoded by the constrained encoder based on SRT (see Section \ref{Sec:Method_SRT}), followed by the single edit reconstruction code encoder (see Section 4.2). After these encoding steps, the unique seed was appended to the beginning of the codeword. Subsequently, a two-bit (i.e. one-nucleotide) coding scheme indicator, which informed the decoder whether Coding Scheme 1 or 2 was in use, was added to both the beginning and end of the oligo, completing the encoding stage. More specifically, we used ’A’ to indicate Coding Scheme 1, and ’C’ for Coding Scheme 2. We remark that this indicator was included solely for marking the two coding schemes to facilitate their performance comparison in our experiment. Since practical data storage systems typically adopt only one coding scheme, this two-bit indicator is not considered when calculating the system’s net information density. 

In the next stage, the binary sequence produced in stage one was converted into a quaternary DNA base sequence, which was then screened by the homopolymer run length and GC-content constraints. Only sequences with a homopolymer run length of no more than 3 and a GC-content between 45\% and 55\% were considered valid and included in the oligo design file. Key parameters of each encoding scheme are illustrated below.

\begin{itemize}
    \item Modified-R10 code encoding: In our experiment, the binary input data file was divided into segments of lengths 452 and 466 bits, for Coding Schemes 1 and 2, respectively. 
    Similar to \cite{erlich2017dna}, each encoded symbol consists of two parts: a unique seed and a data payload. Our seed is of 8 nucleotides long. In Coding Schemes 1 and 2, 28,500 and 28,970 source symbols were encoded into 30,000 symbols, leading to the outer ECC code rates of 0.95 and 0.9657, respectively.
    \item Modified-SRT encoding: 5 nucleotides were introduced as redundancy to forbid homopolymer long runs. In this study, we limited the maximum run length to 3, hence removed all the homopolymer runs of length 4 or more.
    \item Single edit reconstruction code encoding: 9 and 2 nucleotides were added as redundancy to tackle the potential edit errors (insertions, deletions, or substitution errors) introduced by the DNA channel in Coding Schemes 1 and 2, respectively. Since Illumina sequencing can achieve raw error rate of $5 \times 10^{-3}$, an inner ECC that can correct a single edit error within a codeword is adopted by this work.

\end{itemize}

Following the encoding schemes illustrated above, the information densities for Coding Scheme 1 and Coding Scheme 2 are $2 \times \frac{226}{248} \times \frac{28500}{30000} = 1.731$ bits/nt and $2 \times \frac{233}{248} \times \frac{28970}{30000} = 1.815$ bits/nt, respectively.

\subsection{Experimente Design}

The oligos pool was synthesized by Twist Bioscience, incorporating the primer 'GCAGCCCAATGTGTTCGGTCTAC' on the $5^{\prime}$ side and 'ACTGGGTGTTGTCTCTTCGAGCC' on the $3^{\prime}$ side. There were 30,000 oligos synthesized for Coding Schemes 1 and 2, respectively.

After receiving the synthesized oligos pool and amplifying it through PCR, the samples were sequenced by NovogeneAIT using the Illumina HiSeq. As at most 250 nucleotides can be sequenced in a single read for the Illumina sequencer, each oligo was sequenced twice, starting from one end of the oligo and moving towards the other end, leading to the forward and backward fragments of each oligo. These fragments were then merged utilizing the Fast Length Adjustment of SHort reads (FLASH) algorithm \cite{magovc2011flash}. Following the data filtration, 24 million oligos were obtained finally. As shown in Figure \ref{fig:copy all}, the mean coverage was 404.0, with a standard derivation of 234.8.

\subsection{Raw Data Analysis}
Based on the FASTQ file generated by the Illumina sequencer, we evaluated the base error rate, oligo error rate, sequenced oligo length distribution, as well as the read number distribution. These evaluations and analyses were performed on all the oligos rather than for each coding scheme separately, since the distributions of both coding schemes showed significant similarity. 

\begin{figure}[!t]
    \centering
    \includegraphics[width=1\linewidth]{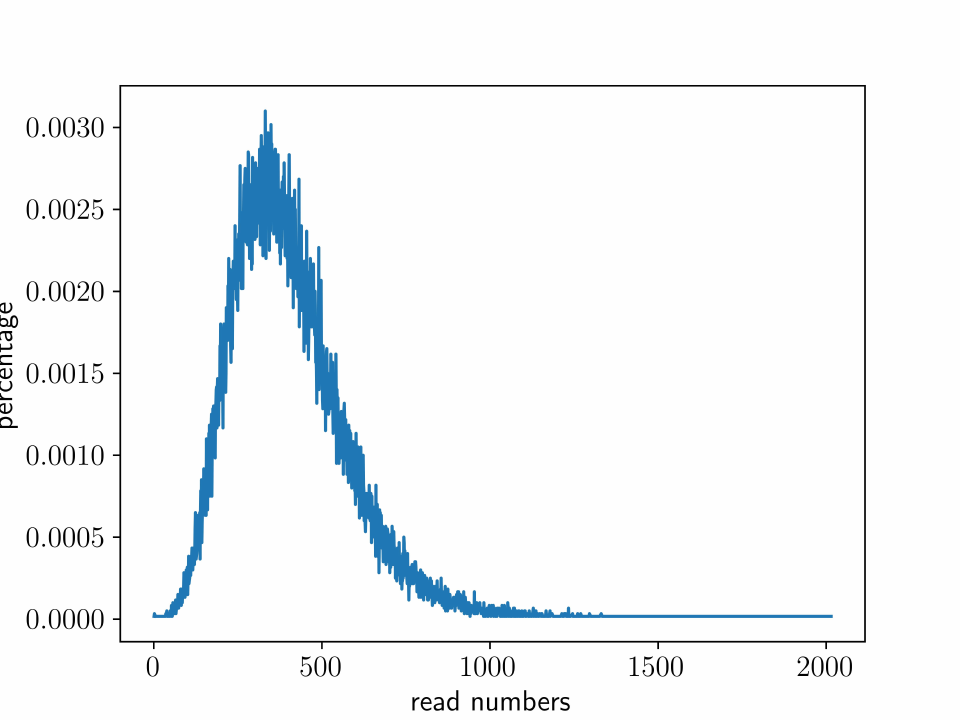}
    \caption{Read number distribution of all the oligos.}
    \label{fig:copy all}
\end{figure}

\begin{figure}[!t]
    \centering
    \includegraphics[width=1\linewidth]{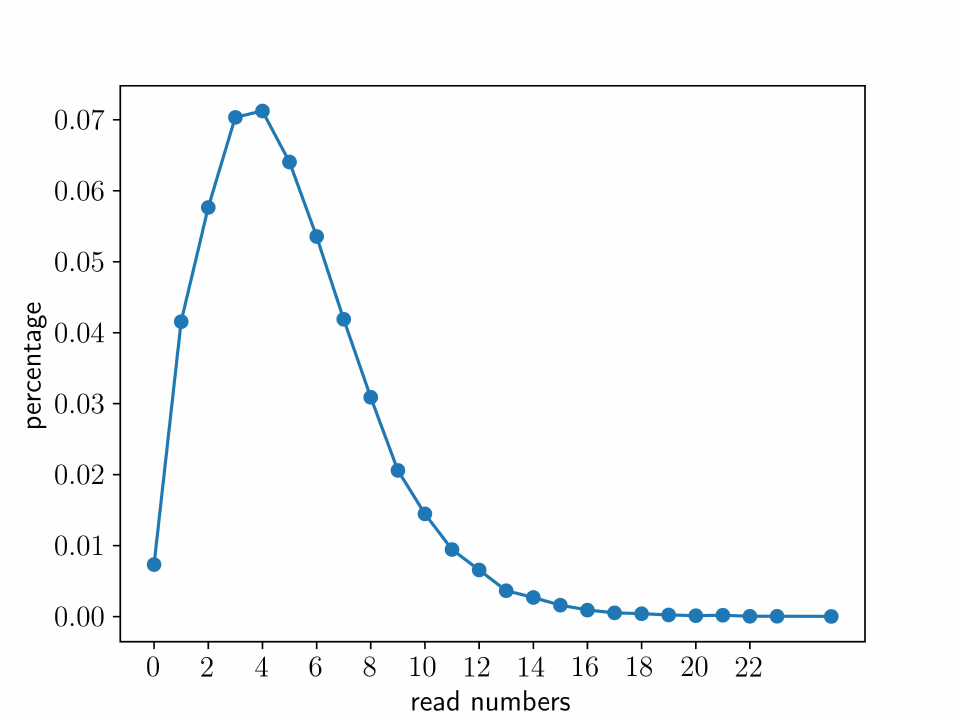}
    \caption{Read number distribution of coverage equals 5x.}
    \label{fig:copy 5x}
\end{figure}

The distributions of read number for all oligos and those with coverage of 5x are illustrated in Figure \ref{fig:copy all} and Figure \ref{fig:copy 5x}, respectively. Furthermore, we also analyzed the base error rate within the data payload region and the results are presented in Table \ref{tab:baseErrorRate}. This analysis reveals that substitutions are the most dominant type of error, which are about 2.5 times as much as deletion errors, and 14.5 times as insertion errors. Our observations are consistent with those reported in \cite{organick2018random}. 

The analysis of the oligo error rate is shown in Table \ref{tab:stringErrorRate}. The length distribution of all sequenced oligos is depicted in Table \ref{tab:length}. Both analyses indicate that majority of the sequenced oligos are of length 250 nts. Furthermore, the number of oligos with lengths less than 250 nts are much larger than that with lengths more than 250 nts. This is due to the much higher deletion error rate than the insertion error rate, as shown by Table \ref{tab:baseErrorRate}.

\begin{table}[!t]
    \centering
    \caption{Error rate of nucleotides except for primer region after sequencing.}
    \resizebox{\linewidth}{!}{
    \begin{tabular}{|c|c|c|c|c|}
       \hline
        & insertion & deletion & substitution & total rate \\
       & rate & rate & rate & /\%\\       
       \hline
       A & $9.21 \times 10^{-6}$ & $6.44 \times 10^{-5}$ & $1.16 \times 10^{-4}$ & 24.67\\
       \hline
       T & $4.32 \times 10^{-6}$ & $6.36 \times 10^{-5}$ & $1.20 \times 10^{-4}$ & 25.39\\
       \hline
       C & $5.81 \times 10^{-6}$ & $6.50 \times 10^{-5}$ & $1.72 \times 10^{-4}$ & 25.48\\
       \hline
       G & $2.37 \times 10^{-5}$ & $6.34 \times 10^{-5}$ & $2.17 \times 10^{-4}$ & 24.47\\
       \hline
       Total & $4.31 \times 10^{-5}$ & $2.56 \times 10^{-4}$ & $6.25 \times 10^{-4}$ & 100\\
       \hline
    \end{tabular}
    }
    
    \label{tab:baseErrorRate}
\end{table}

\begin{table}[!t]
    \centering
    \caption{Oligo error rates due to various types of errors.}
    \begin{tabular}{|c|c|c|c|c|c|}
        \hline
       \makecell[c]{edit \\ errors}  & ins & del & sub & numbers & rate/\% \\
       \hline
        0 & 0 & 0 & 0 & 20380197 & 85.77 \\
        \hline
        1 & 0 & 0 & 1 & 1932524 & 8.133 \\
        \hline
        1 & 0 & 1 & 0 & 440656 & 1.854 \\
        \hline
        1 & 1 & 0 & 0 & 214364 & 0.902 \\
        \hline
        2 & 0 & 0 & 2 & 224483 & 0.945 \\
        \hline
        2 & 0 & 2 & 0 & 83018 & 0.349 \\
        \hline
        2 & 0 & 1 & 1 & 41528 & 0.175 \\
        \hline
        2 & 1 & 0 & 1 & 20554 & 0.0865 \\
        \hline
        2 & 1 & 1 & 0 & 4568 & 0.0192 \\
        \hline
        2 & 2 & 0 & 0 & 1328 & 0.00559 \\
        \hline
        $>2$ & - & - & - & 419272 & 1.764 \\
        \hline
    \end{tabular}
    \label{tab:stringErrorRate}
\end{table}

\begin{table}[!t]
    \centering
    \caption{Sequenced oligo length distribution.}
    \begin{tabular}{|c|c|c|}        
        \hline
        length/nts & numbers & rate/\% \\
        \hline
        $<249$ & 507798 & 2.09\\
        \hline
        249 & 536876 & 2.21\\
        \hline
        250 & 22936606 & 94.62\\
        \hline
        251 & 255939 & 1.06\\
        \hline
        $>251$ & 7827 & 0.032\\
        \hline
    \end{tabular}
    \label{tab:length}
\end{table}

\subsection{Data Recovery}
\begin{figure*}[!t]
    \centering
    \includegraphics[width=1\linewidth]{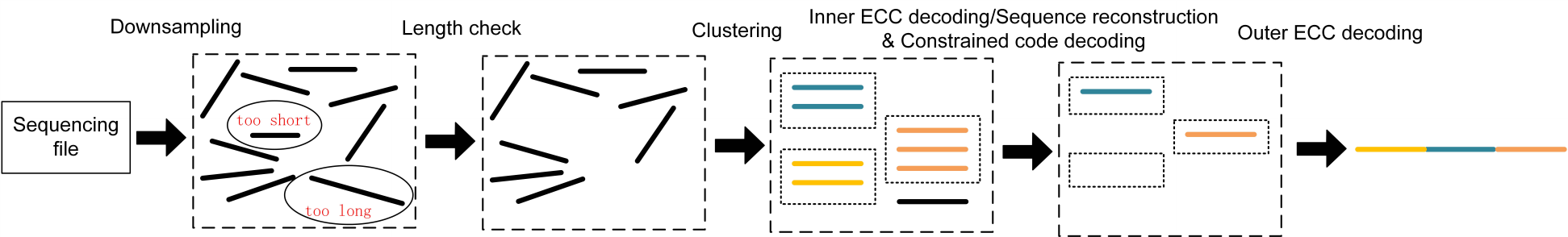}
    \caption{The workflow of the downsampling and data recovery experiment. After downsampling, length check was conducted to discard oligos of lengths less than 249 nts or more than 251 nts. Next, the oligos were clustered based on their seeds. The sequence reconstruction was implementing on the sequences sharing the same seed. Thereafter, the constrained code decoding was applied on each sequence, following by the outer ECC decoding to recover the originally stored user data.}
    \label{fig:decoding flow}
\end{figure*}

For a given coding scheme and coverage $N$, the process of downsampling is shown in Figure \ref{fig:decoding flow}. Specifically, the downsampling is first carried out by randomly taking $30000 \times N$ sequences from the corresponding 12 million sequenced oligos. Next, oligos whose lengths exceed the range of [249, 251] were discarded as the inner ECC adopted by this experiment can only correct one edit error. Subsequently, oligos were clustered by their seeds, followed by the logic decoding. The logic decoding comprises three sequential steps: Sequence reconstruction (inner ECC decoding), followed by constrained code decoding, and outer ECC decoding.

More specifically, the sequence reconstruction was performed on each cluster, and the whole cluster will be discarded if the sequence reconstruction result does not form a codeword. Two different inner ECC decoding approaches were applied, namely the ``detection" approach and the ``decoding" approach. For sequences that do not form valid codewords, the ``detection" approach discarded them whereas the ``decoding" approach performed decoding on the sequence.
Subsequently, constrained code decoding was conducted, where the invalid sequences were also discarded (see Section 4.1). 
Finally, the outer ECC decoding was performed on all the available sequences using the modified basis-finding algorithm (modified-BFA, see Section 4.5). The BFA \cite{he2023basis} was designed for decoding fountain code for channels where the received symbols could have errors.


The downsampling and data recovery experiment completed with the outer ECC decoding. It was considered successful only when the original information sequence is fully recovered without any error. Assuming $T$ number of downsampling and data recovery experiments are made and $S$ of them are successful, we define the successful data recovery rate as $S/T$. The successful data recovery rate for different coding schemes and coverage numbers is depicted in Figure \ref{fig:FER}. At each coverage, the success rate was calculated based on downsampling and data recovery experiments conducted 1,000 times. An upper bound for the outer ECC decoder can also be obtained by performing the the maximum likelihood (ML, i.e., optimal) decoding exclusively on the correctly received symbols (see Supplementary Material Section 1.2). It is labeled  as `outer ECC UB' in Figure \ref{fig:FER}. From Figure \ref{fig:FER}, we observe that:


\begin{figure*}[!t]
    \centering
    \begin{subfigure}[b]{0.48\linewidth}
        \includegraphics[width=\linewidth]{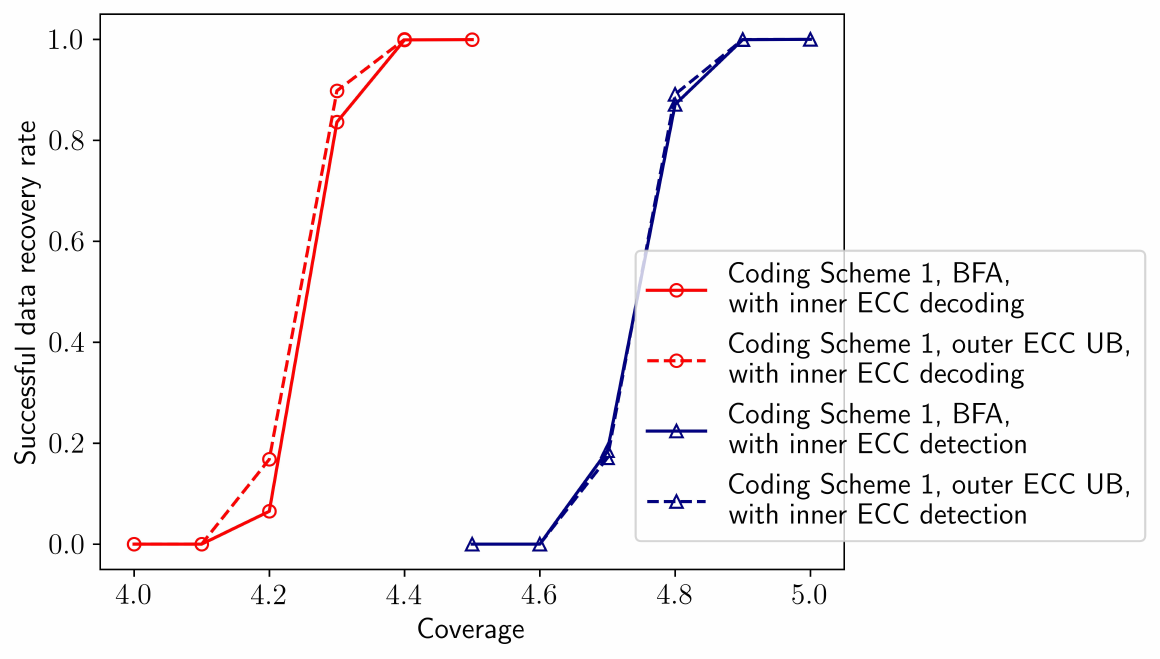}
        \caption{Decoding performance of Coding Scheme 1.}
        \label{fig:FER_scheme1}
    \end{subfigure}
    \hfill
    \begin{subfigure}[b]{0.48\linewidth}
        \includegraphics[width=\linewidth]{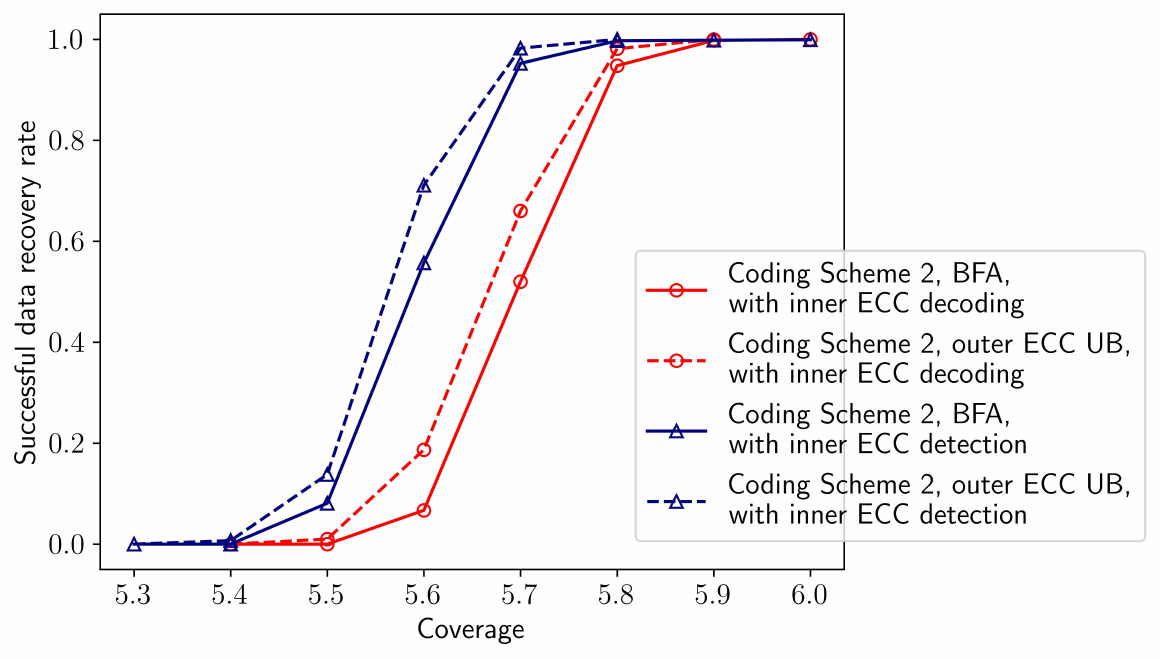}
        \caption{Decoding performance of Coding Scheme 2.}
        \label{fig:FER_scheme2}
    \end{subfigure}
    \caption{Successful data recovery rate among 1,000 experiments, with different sequence reconstruction approaches for Coding Schemes 1 and 2. The ``outer ECC UB" represents the theoretical best performance of outer decoder given inner decoding method.}
    \label{fig:FER}
\end{figure*}

\begin{itemize}
    \item Compared to Coding Scheme 2, Coding Scheme 1 had a higher successful data recovery rate with the same sequencing coverage, due to the higher coding redundancy introduced to the coded DNA strands.
    \item The modified-BFA approached the theoretical best performance with both inner ECC decoding approaches, demonstrating its effectiveness for decoding fountain codes for both Coding Schemes 1 and 2. This is because BFA was specifically designed for decoding fountain codes for channels whose received symbols could be corrupted by errors.
    
    \item For Coding Scheme 1, the decoding approach of inner ECC led to better performance UB than the detection approach, while the opposite trend was observed for Coding Scheme 2. This is due to the reduced redundancy of Coding Scheme 2. As a result, if an incorrect codeword is decoded by the inner ECC decoder, a large number of candidate codewords will be generated during the outer ECC decoding, and most of them will be incorrect, leading to more decoding errors. For example, decoding of a length $l$ sequence with one substitution will generate $l$ codewords, and only one of them is correct.
    
\end{itemize}


\section{Discussion}

In this work, we encoded 1.61 MB and 1.69 MB of data into 30,000 oligos each, achieving net information densities of 1.731 and 1.815, respectively. All information was fully recovered without errors at sequencing coverages of 4.5 and 6.0. This experiment achieved the highest net information density and lowest coverage among existing DNA storage experiments (with standard DNA), with data recovery rates and coverage approaching theoretical optima. The superior performance was mainly benefited from the following coding algorithms and techniques.

\begin{itemize}
    \item In this work, constrained encoding was carried out in two stages to ensure that the synthesized oligos satisfy the constraints of a homopolymer runlength of 3 and a GC content between 45\% and 55\%. In stage 1, since the homopolymer runlength constraint is a strong deterministic constraint, the constrained encoding was conducted based on SRT. In stage 2, the GC-content constraint was imposed through the screening process on a set of candidate oligos generated by the Raptor encoder using different seeds. As a result, a total of 11 redundant bits were introduced to impose the two biological constraints, leading to a constrained code rate of 1.9556 and coding efficiency of 98.69\%. Furthermore, compared with the DNA Fountain code \cite{erlich2017dna}, the encoding latency is also significantly reduced (see Supplementary Material, Section 4).
    \item Since Illumina sequencing can achieve raw error rate of $5 \times 10^{-3}$, single edit reconstruction codes that are asymptotically optimal \cite{cai2021coding} were adopted as the inner ECC, with redundancy of 9 nts for Coding Scheme 1 and 2 nts for Coding Scheme 2. The corresponding reconstruction codes decoding algorithms proposed in \cite{cai2021coding} can effectively utilize the information provided by all the oligos sharing the same seed to recover the original source oligo.
    \item In this work, we adopted the Raptor code as the outer ECC, which enhances the LT code by adding a pre-code, enabling linear encoding/decoding complexity and improving error correction. Moreover,  in DNA storage channels, limited redundancy in the inner ECC may not  adequately detect or correct all the errors, leading to undetectable errors at the output of the inner ECC decoder. The state-of-the-art decoders of fountain codes cannot address these undetectable errors. Therefore, we developed a modified basis-finding algorithm (modified-BFA) to decode the fountain codes. Other than the oligo lost, it can also mitigate the erroneously received sequences. Our modified BFA facilitates higher net information density and lower sequencing coverage, such it can effectively correct the undetectable errors of the inner codes, including both the inner ECC and constrained codes. For example, with SRT, which has low redundancy, constrained decoding may lead to significant error propagation. However, because all errors are confined within a DNA strand (i.e., a symbol of the fountain code), they can be effectively mitigated by the modified-BFA decoding.
\end{itemize}

\section{Methods}

\subsection{Modified-SRT}

\label{Sec:Method_SRT}

\begin{figure}
    \centering
    \includegraphics[width=1\linewidth]{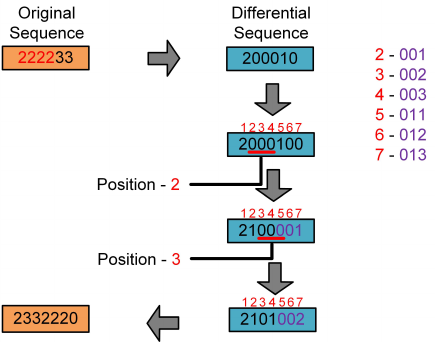}
    \caption{An example of encoding by SRT to obtain a sequence satisfying the run length constraint of 3.}
    \label{fig:SRT example}
\end{figure}

The SRT has been widely applied \cite{immink2017design, nguyen2021capacity} to efficiently remove the forbidden substrings within a codeword, such as the homopolymer runs. In this work, we applied the algorithm presented in \cite{nguyen2021capacity}, which can convert any given quaternary sequence of length $n < 50$ to a sequence of length $n + 1$ satisfying the run length constraint of 3. The key step is to transform the sequence into a so-called differential sequence. For a sequence $\mathbf{x} = (x_1, x_2, ..., x_n) \in \{0, 1, 2, 3\} ^ n$, the differential sequence $\mathbf{y} = (y_1, y_2, ... y_n) \in \{0, 1, 2, 3\} ^ n$ is defined by $y_1 = x_1$ and $y_i = x_i \oplus x_{i - 1}$ ($\oplus$ represents the bitwise XOR operation) for $2 \leq i \leq n$. In \cite{nguyen2021capacity}, the differential sequence is derived using the module 4 subtract operation, whereas we used the bitwise XOR operation in our experiments, since it is more hardware-friendly. The encoding process is illustrated as follows:

\begin{itemize}
    \item Step 1: Compute the differential sequence $\mathbf{y}$ from the original sequence $\mathbf{x}$.
    \item Step 2: Append a '0' to the end of $\mathbf{y}$.
    \item Step 3: Starting with ${y}_2$, search for and remove sequences of 3 consecutive zeros in $\mathbf{y}$, and append their positions to the end of $\mathbf{y}$ until no such sequence remain, yielding a sequence $\mathbf{y}^{\prime}$. In this step, to differentiate with the adding `0' in Step 2, all the positions do not end with zero.
    \item Step 4: Output the sequence whose differential sequence is $\mathbf{y}^{\prime}$.
\end{itemize}
We provide a toy example illustrating the encoding process by Figure \ref{fig:SRT example}. This example shows how to encode the sequence $\mathbf{x} = 222233$ utilizing SRT.

To illustrate why removing all sequences of 3 consecutive zeros in the differential sequence can obtain a sequence satisfying the homopolymer run length constraint of 3, consider the fact that each zero in the differential sequence, except for the first position, corresponds to two consecutive identical symbols in the original sequence. Therefore, a consecutive zeros of length $l$ in the differential sequence indicates a consecutive identical symbols of length $l + 1$ in the original sequence. 

\begin{figure}[!t]
    \centering
    \includegraphics[width=1\linewidth]{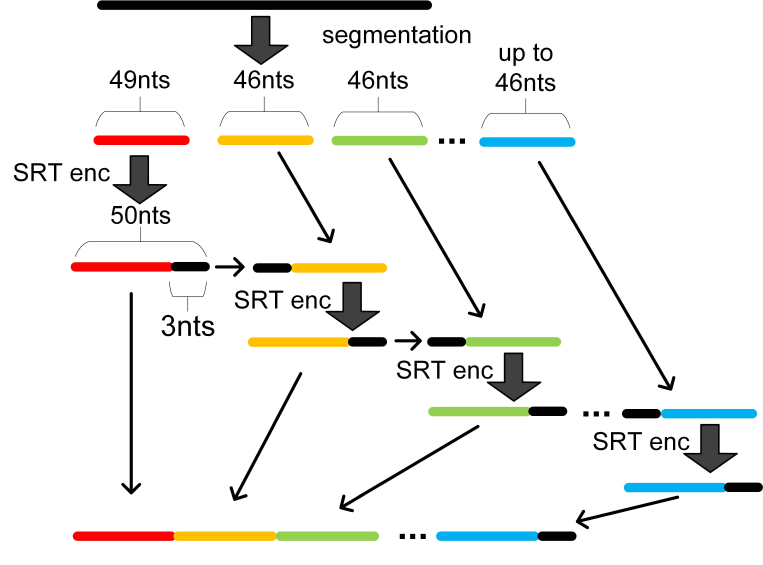}
    \caption{The encoding process of modified-SRT for run length constraint of 3. Firstly, the input sequence is segmented into several data blocks. The first block has a length of 49, while the other blocks have length of 46. Except for the first data block, each data block takes the last 3 symbols of the previous encoded block as its beginning symbols before conducting SRT encoding.}    
    \label{fig:SRT encoding}
\end{figure}

To ensure fixed-length encoding, the position indicator added in is always 3 bits long. Furthermore, to distinguish with the 0 added in the Step 2, the position cannot end up with a bit zero. These requirements confine the block size to a maximum of $4 \times 4 \times 3 + 1 = 49$ for the run length constraint of 3.
In this experiment, the input sequences of the SRT encoder have lengths of 234 and 241 nucleotides, for Coding Schemes 1 and 2, respectively, forbidding the direct application of the encoding scheme described above (requiring input length $n < 50$). Therefore, in this work, as illustrated by Figure \ref{fig:SRT encoding}, we modified the SRT to enable the encoding of sequences with arbitrary length. In particular, the last three symbols of the current encoded block are concatenated with the subsequent data block before encoding it, ensuring that the homopolymer run length constraint is satisfied at the boundaries of the encoded blocks. The decoding is simply the reverse of the encoding.

\subsection{Single edit reconstruction code}
\label{Sec:Method_Reconstruction}

In Coding Scheme 1, 9 nucleotides were introduced as redundancy. Define $P_1 = \{1, 2, 4, 8, 16, 32, 64, 128, 248\}$. Let $\mathbf{y} = \mathrm{enc}_1(\mathbf{x})$ denote an encoder that is able to encode the input binary string $\mathbf{x} = (x_1, x_2, ... , x_{239}) \in \{0, 1\} ^ {239}$ into output string $\mathbf{y} = (y_1, y_2, ... , y_{248}) \in \{0, 1\} ^ {248}$ satisfying $\sum_{1 \leq i \leq 248} i \cdot y_i \equiv 0 \pmod{496}$. The $y_p$'s, $\forall p \in P_1$ serve as redundancy, and the other bits in $\mathbf{y}$ are information bits taken directly from $\mathbf{x}$. During encoding, the input sequence $\mathbf{x} = (x_1, x_2, ... , x_{478}) \in \{0, 1\} ^ {478}$ is segmented into odd and even indexed sequences $\mathbf{x}_o = (x_1, x_3, ..., x_{477})$ and $\mathbf{x}_e = (x_2, x_4, ..., x_{478})$, respectively. They are then sent to $\mathrm{enc}_1$ to generate $\mathbf{y}_o$ and $\mathbf{y}_e$, respectively. Subsequently, $\mathbf{y}_o$ and $\mathbf{y}_e$ are re-combined to form the complete encoded sequence.


In Coding Scheme 2, 2 nucleotides were added as redundancy. Similar to Coding Scheme 1, denote $P_2 = \{247, 248\}$ and let $\mathbf{y} = \mathrm{enc}_2(\mathbf{x})$ represent an encoder than encodes the quaternary input string $\mathbf{x} = (x_1, x_2, ..., x_{246}) \in \{0, 1, 2, 3\} ^ {246}$ into coded sequence $\mathbf{y} = (y_1, y_2, ..., y_{248}) \in \{0, 1, 2, 3\} ^ {248}$ satisfying $\oplus_{1 \leq i \leq 248, i \equiv 0 \pmod{2}}y_i = 0$ and $\oplus_{1 \leq j \leq 248, j \equiv 1 \pmod{2}}y_j = 0$. The $y_p$'s, $\forall p \in P_2$ serve as redundancy, while the other bits in $\mathbf{y}$ are information bits taken directly from $\mathbf{x}$.

For both Coding Schemes 1 and 2, only received sequences with zero syndromes are considered valid codewords. To recover the received sequence with one edit error, the decoder first finds out all the potential codewords within one edit distance from the received sequence in an exhaustive manner. Assuming the received sequence $\mathbf{y}$ is not a valid codeword, and it has length $m^{\prime}$ while the original codeword has length $n^{\prime}$, there are three scenario: (1) If $m^{\prime} = n^{\prime} + 1$, each nucleotide in $\mathbf{y}$ is deleted one by one to check whether the resulting sequence forms a valid codeword. (2) If $m^{\prime} = n^{\prime}$, each nucleotide in $\mathbf{y}$ is replaced by the other three types of nucleotides to check whether the resulting sequence forms a valid codeword. (3) If $m^{\prime} = n^{\prime} - 1$, a nucleotide is inserted in each position in $\mathbf{y}$ to check whether the resulting sequence forms a valid codeword. If $m^{\prime} < n^{\prime} - 1$ or $m^{\prime} > n^{\prime} + 1$, it is obvious that more than one edit errors occur, which are beyond the error-correction capability of the single-edit correcting codes.

For sequence reconstruction with more reads, as proposed by \cite{cai2021coding}, a list of valid codewords within one edit distance is generated, for each received sequence within a cluster. In particular, if the received sequence is a valid codeword, the list contains only the sequence itself. Otherwise, it includes all codewords with one edit distance from the received sequence, as illustrated above. Note that the list may be empty. Thereafter, the unique codeword that appears most frequently across all lists of codewords is selected as the decoded sequence, and the corresponding frequency is used for the reliability ranking during outer ECC decoding (see Section \ref{Sec:Method_BFA}). If no such codeword is found, the decoding result is empty.



\subsection{Modified-R10 Codes}
\label{Sec:R10}
 
The fountain codes, originally designed for the erasure channel, where an encoded symbol transmitted over the channel is either lost or received with no error, exhibit a rateless property. Therefore, fountain codes can be efficiently combined with constrained codes, by examining the code constraints on each fountain codeword through a screening process. Thus, prior work \cite{erlich2017dna} employed LT codes \cite{luby2002lt}, the first practical realization of fountain codes, as the outer ECC to achieve a high information storage density. Another class of fountain codes, Raptor codes \cite{shokrollahi2006raptor}, can achieve better error-correction than the LT codes while remaining linear encoding and decoding by incorporating a pre-code.

In our experiments, we developed a modified-R10 encoding scheme. First, the pre-code encoder, which is a $(n_p, k_p)$ linear block encoder, transforms the $k_p$ source symbols into $n_p$ intermediate symbols. Let $\mathbf{D}$ be the intermediate symbols matrix, each row of which represents an intermediate symbol of bit-width $l_d$, and let $\mathbf{H}_p$ denote the parity-check matrix of the pre-code, which is designed in advance and known to both the sender and receiver. $\mathbf{D}$ and $\mathbf{H}_p$ are constrained by the pre-code encoding to satisfy $\mathbf{H}_p \cdot \mathbf{D} = \mathbf{0}$, where $\mathbf{0}$ denote the all-zero matrix. It is then followed by the LT code encoding. 


The LT code encoding is carried out for multiple rounds, each of which outputs an encoded symbol, until a sufficient number of encoded symbols are obtained.
At the beginning of each LT code encoding round, a seed $s$ is generated and used to initialize a PRNG. The PRNG then outputs a length-$n_p$ vector $\mathbf{g}$, which we call generator vector. In particular, $\mathbf{g}$ consists of $d$ ones and $n_p-d$ zeros, where the integer $d$ and the positions of the $d$ ones are uniquely determined by the seed $s$ and the PRNG. Thereafter, a so-called data payload vector $\mathbf{d}$ is derived by $\mathbf{d} = \mathbf{g} \cdot \mathbf{D}$. The encoded symbol is then formed as the seed $s$ concatenated with the data payload vector $\mathbf{d}$. 

The modified-R10 code differs from the R10 code in two aspects. First, the R10 code allows a maximum of 8,192 distinct seeds, whereas in the modified-R10 code, the seed is initialized to zero and then increased by one in each encoding round until the encoder output reaches 30,000 oligos for both coding schemes. The second modification is that the PRNG is implemented using the Mersenne Twister \cite{matsumoto1998mersenne}, which generates numbers with improved randomness compared to the standard PRNG.



\subsection{Modified-BFA}
\label{Sec:Method_BFA}

In prior work [6], the belief propagation (BP) decoding was employed to decode the fountain code (i.e. the LT code). However, the BP algorithm has significantly worse performance for decoding Raptor codes due to pre-coding. 
The Structured Gaussian Elimination (SGE) algorithm, also known as inactivation decoding, is considered the maximum likelihood (ML, i.e., optimal) decoder for decoding fountain codes over the erasure channels \cite{shokrollahi2011raptor}.
However, in DNA-based data storage channels, limited redundancy in the inner ECC may not adequately detect or correct all the errors, leading to undetectable errors at the output of the inner ECC decoder. Both SGE and BP algorithms cannot address these undetectable errors. Therefore, in this work, we developed a modified-BFA to decode the Raptor codes. Other than the oligo lost, it can also mitigate the erroneously sequences generated by the inner ECC decoder.

First, the modified-BFA decoder initializes a received matrix $\mathbf{R} = [\mathbf{H}_p, 0]$, where $\mathbf{0}$ represents the all-zero matrix with size $(n_p - k_p) \times l_d$ (see Section \ref{Sec:R10}). 
Next, all sequences generated by the inner ECC decoder, each of which corresponds to an encoded symbol of the modified-R10 code, are sorted by their appearance frequencies during inner ECC decoding (see Section \ref{Sec:Method_Reconstruction}) from the highest to the lowest.
These sequences are processed one-by-one in order. Specifically, each one is converted into a length-$(n_p+l_d)$ vector $\mathbf{r}$, where the first $n_p$ bits and last $l_d$ bits respectively correspond to the generator vector created by the seed and the data payload vector in the sequence, and $\mathbf{r}$ is then appended to the end of $\mathbf{R}$.
In this case, the rows of $\mathbf{R}$ have decreasing reliability from the top to the bottom.

Thereafter, the standard BFA \cite{he2023basis} is performed on $\mathbf{R}$ with three steps. First, the decoder finds a basis $B$ of $\mathbf{R}$, i.e., $B$ is a minimum subset of $\mathbf{R}$'s rows such that $B$ is able to linearly represent all $\mathbf{R}$'s rows. Specifically, $B$ is initialized as an empty set. The decoder then traverse the rows of $\mathbf{R}$ from top to bottom and a row is added into $B$ if it cannot be linearly represented by rows in $B$. Subsequently, the decoder computes the frequency that each basis vector in $B$ attends the linear representation of the rows of $\mathbf{R}$. Finally, SGE is performed on the basis vectors that are most frequently involved in the linear representations.

\bibliographystyle{naturemag}
\bibliography{myreference}

\end{document}